# OntoELAN: an Ontology-based Linguistic Multimedia Annotator

Artem Chebotko, Yu Deng, Shiyong Lu, Farshad Fotouhi
*Wayne State University*
*Department of Computer Science*
*5143 Cass Avenue, Detroit, Michigan 48202, USA*
*Contact author e-mail: artem@cs.wayne.edu*

Anthony Aristar
*Wayne State University*
*Department of English*
*51 W.Warren, Detroit, Michigan 48202, USA*

Hennie Brugman, Alexander Klassmann, Han Sloetjes, Albert Russel, Peter Wittenburg
*Max Planck Institute for Psycholinguistics*
*Nijmegen, The Netherlands*

## Abstract

*Despite its scientific, political, and practical value, comprehensive information about human languages, in all their variety and complexity, is not readily obtainable and searchable. One reason is that many language data are collected as audio and video recordings which imposes a challenge to document indexing and retrieval. Annotation of multimedia data provides an opportunity for making the semantics explicit and facilitates the searching of multimedia documents. We have developed OntoELAN, an ontology-based linguistic multimedia annotator that features: (1) support for loading and displaying ontologies specified in OWL; (2) creation of a language profile, which allows a user to choose a subset of terms from an ontology and conveniently rename them if needed; (3) creation of ontological tiers, which can be annotated with profile terms and, therefore, corresponding ontological terms; and (4) saving annotations in the XML format as Multimedia Ontology class instances and, linked to them, class instances of other ontologies used in ontological tiers. To our best knowledge, OntoELAN is the first audio/video annotation tool in linguistic domain that provides support for ontology-based annotation.*

**Keywords**: ontology, annotation, multimedia, Semantic Web, OWL, GOLD.

**Acknowledgements**: We would like to thank Dr. Laura Buszard-Welcher and Andrea Berez from E-MELD (Electronic Metastructure for Endangered Languages Data) project for their constructive comments on *OntoELAN*.

## 1. Introduction

Many languages are in serious danger of being lost and if nothing is done to prevent it, half of the world's approximately 6,500 languages will disappear in the next 100 years. Language data are central to the research of a large social science community, including linguists, anthropologists, archeologists, historians, sociologists, and political scientists interested in the culture of indigenous people. The death of a language entails the loss of a community's traditional culture, for the language is a unique vehicle for its traditions and culture. Recently, there is an increasing interest and effort for preserving and documenting endangered languages [12, 2].

Despite its scientific, political, and practical value, comprehensive information about human languages, in all their variety and complexity, is not readily obtainable and searchable. One reason is that many language data are collected as audio and video recordings which imposes a challenge to document indexing and retrieval. Annotation of multimedia data provides an opportunity for making the semantics explicit and facilitates the searching of multimedia documents. However, different annotators might use different vocabulary to annotate multimedia, which cause low recall and precision in search and retrieval. In this paper, we propose an ontology-based annotation approach, in which a linguistic ontology is used so that the terms and their relation-ships are formally defined. In this way, annotators will use the same vocabulary to annotate multimedia, and search engines driven by the ontology will retrieve multimedia data with greater recall and precision.

We present an ontology-based linguistic multimedia annotation tool *OntoELAN* – a successor of EUDICO Linguistic Annotator (*ELAN*) [10] developed at the Max Planck Institute for Psycholinguistics, Nijmegen, The Netherlands, with the aim to provide a sound technological basis for the annotation and exploitation of multimedia recordings. Although *ELAN* is designed specifically for linguistic domain (analysis of language, sign language, and gesture), it can be used for annotation, analysis, and documentation purposes in other multimedia domains. We briefly describe the features of *ELAN* later in this paper and refer the reader to [10] for details.

*OntoELAN* inherits all *ELAN*'s features and extends the tool with an ontology-based annotation approach. In particular, our main contributions are:

- *OntoELAN* can open and display ontologies, specified in OWL Web Ontology Language [5];

- *OntoELAN* allows the creation of a language profile, which allows a user to choose a subset of terms from an ontology and conveniently rename them if needed.

- *OntoELAN* allows the creation of ontological tiers, which can be annotated with profile terms and, therefore, corresponding ontological terms.

- *OntoELAN* saves annotations in XML [7] format as class instances of Multimedia Ontology, which is designed based on XML Schema [8] of *ELAN* annotation files.

- *OntoELAN*, while annotating ontological tiers, creates class instances of corresponding ontologies linked to annotation tiers and relates them to instances of Multimedia Ontology classes.

*OntoELAN* is developed to fulfill annotation requirements for linguistic domain. It is natural, that, in this paper, we use linguistic annotation examples and General Ontology for Linguistic Description (GOLD) [9] linked to an ontological tier. To our best knowledge, *OntoELAN* is the first audio/video annotation tool in linguistic domain that provides support for ontology-based annotation.

## 2. Related work

Linguistic domain places some minimum requirements on multimedia annotation tools. While semantics-based contents such as speeches, gestures, signs, people scenes are important, color and texture are not of that interest. To annotate semantics-based content, a tool should provide a time axis and a possibility of its subdivision into time slots/segments, multiple tiers for different semantic content, etc. Obviously, there should be some multimedia resource metadata like title, authors, date and time, etc. Additionally, a tool should provide ontology-based annotation features to enable standard annotations for the whole domain.

As related work, we give a brief description of the following tools: *Protégé* [3], *IBM MPEG-7 Annotation Tool* [1] and *ELAN* [10].

*Protégé* [3] is a popular ontology construction and annotation tool developed at Stanford University. *Protégé* supports the Web Ontology Language through the OWL Plug-in, which allows a user to load OWL ontologies, annotate data and save annotation markup. Unfortunately, *Protégé* provides only simple multimedia support through the Media Slot Widget. The Media Slot Widget allows the inclusion and display of video and audio files in *Protégé*, which may be enough for general description of multimedia files like metadata entries, but not sufficient for annotation of a speech, where the multimedia time axis and its subdivision into segments are crucial.

*IBM MPEG-7 Annotation Tool* [1] was developed by IBM to assist annotating video sequences with MPEG-7 [13] metadata based on the shots of the video. It does not support any ontology language and uses an editable lexicon from which a user can choose keywords to annotate shots. A *shot* is defined as a time period in video in which the frames have similar scenes. Annotations are saved based on MPEG-7 XML Schema [13]. Although the *IBM MPEG-7 Annotation Tool* was specially designed to annotate video, the shot and lexicon based annotation does not provide enough flexibility for linguistic multimedia annotation. In particular, shot approach is good for the annotation of content-based features like color and texture, but not for time alignment and time segmentation required for semantics-based content annotation.

*ELAN* (EUDICO Linguistic Annotator) [10] developed at the Max Planck Institute for Psycholinguistics, Nijmegen, The Netherlands, is designed specifically for linguistic domain (analysis of language, sign language, and gesture) to provide a sound technological basis for the annotation and exploitation of multimedia recordings. *ELAN* provides many important features for linguistic data annotation such as time segmentation, multiple annotation layers, but not the support of an ontology. Annotation files are saved in the XML format based on *ELAN* XML Schema.

As a summary, existing annotation tools such as *Protégé* and *IBM MPEG-7 Annotation Tool* are not suitable for our purpose since they do not support many multimedia annotation operations such as time transcription and translation of linguistic audio and video data. *ELAN* is the best candidate for becoming a widely-accepted linguistic multimedia annotator, and it is already used by linguists throughout

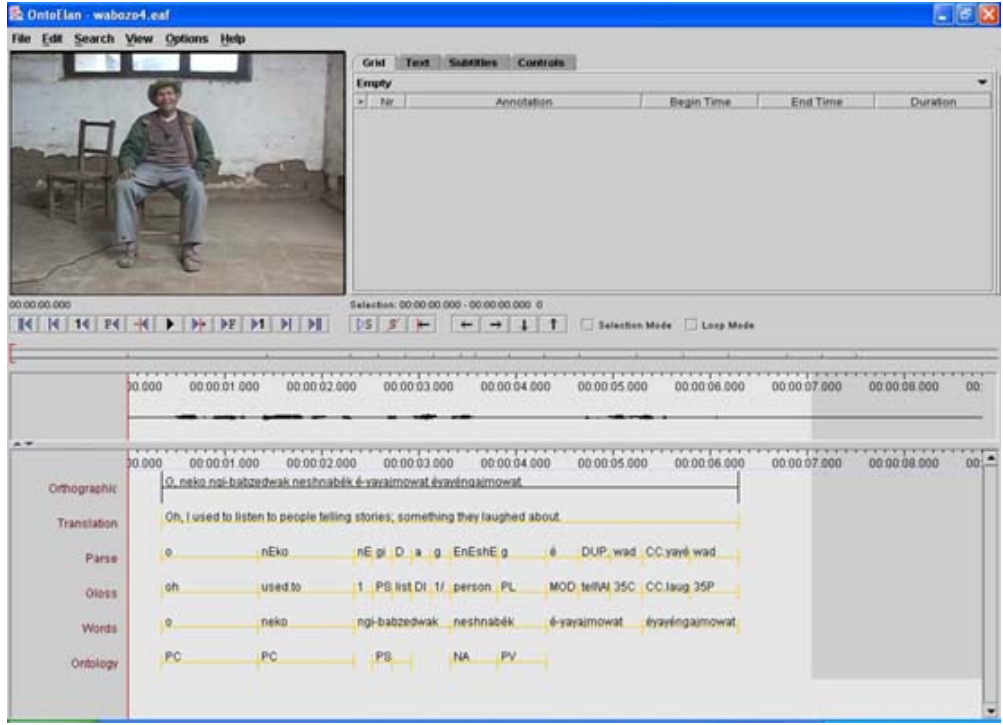

**Figure 1. A snapshot of the OntoELAN main window.**

the world.

## 3. OntoELAN overview

*OntoELAN* is an ontology-based linguistic multimedia annotator, developed on the top of *ELAN* annotator. Currently, *OntoELAN* source code contains more than 60,000 lines of Java code and has several years development history started by Max Planck Institute for Psycholinguistics team and continued by Wayne State University team. Both development teams will continue their collaboration on *ELAN* and *OntoELAN*.

*OntoELAN* has a long list of detailed descriptions of all its technical features including the following features that are inherited from *ELAN*:

- display a speech and/or video signals, together with their annotations;

- time linking of annotations to media streams;

- linking of annotations to other annotations;

- unlimited number of annotation tiers as defined by a user;

- different character sets;

- basic search facilities.

*OntoELAN* implements the following additional features:

- loading of OWL ontologies;

- language profile creation;

- ontology-based annotation;

- storing annotations in XML based on Multimedia Ontology and domain ontologies.

The main window of *OntoELAN* is shown in Figure 1. *OntoELAN* has a video viewer, an annotation density viewer, a waveform viewer, a timeline viewer, and associated with them controls and menus. In this paper we will mostly work with the timeline viewer (see Figure 1), which contains different annotation tiers. Detailed description of the interface is available in [10].

In the following sections, we will focus on the description of features that make *OntoELAN* an ontology-based annotator, like Multimedia Ontology for *OntoELAN* and domain ontologies, a language profile, ontological annotation tiers.

## 4. Ontologies

*OntoELAN* saves its annotations in the XML format as class instances of Multimedia Ontology and class instances of ontologies that are used in ontological tiers. We have developed Multimedia Ontology (available at http://www.cs.wayne.edu/~yudeng/project/elan3/multimedia.owl) to provide a semantic framework for multimedia annotation with *OntoELAN*. It is expressed in Web Ontology Language (OWL) [5] and its design is based on *ELAN* XML Schema for annotation. Multimedia Ontology contains the following classes:

- *AnnotationDocument*, which represents the whole annotation document.

- *Tier*, which represents a single annotation tier/layer. There are several types of tiers that a user can choose.

- *TimeSlot*, which represents a concept of time segment that may subdivide tiers.

- *Annotation*, which can be either *AlignableAnnotation* or *ReferringAnnotation*.

- *AlignableAnnotation*, which links directly to a time slot.

- *ReferringAnnotation*, which can reference an existing *AlignableAnnotation*.

- *AnnotationValue*, which has two subclasses *StringAnnotation* and *OntologyAnnotation* that represents two different ways of annotating.

- *MediaDescriptor*, *TimeUnit* and others.

Among our contributions is the introduction of the OWL class *OntologyAnnotation*, which serves as an annotation unit for an ontology-based annotation. *OntologyAnnotation* has restrictions on the following properties:

- hasOntAnnotationId – the ID of the annotation.

- hasUserDefinedTerm, which relates *OntologyAnnotation* to a term in a language profile (described in the next section).

- hasInstances, which relates *OntologyAnnotation* to a term (represented as an instance) in an ontology used for annotation.

- hasOntAnnotationDescription – descriptions/comments on the annotation.

In general, *AnnotationDocument* may have zero or many *Tiers*, which, in turn, may have zero or many *Annotations*. *Annotation* can be either *AlignableAnnotation* or *ReferringAnnotation*, where *AlignableAnnotation* can be divided by *TimeSlots*, and *ReferringAnnotation* can refer to another annotation. *ReferringAnnotation* may refer to *AlignableAnnotation* as well as to *ReferringAnnotation*, but the root of the referenced annotations must be an *AlignableAnnotation*. Each *Annotation* has one *AnnotationValue*, which can be either a *StringAnnotation* or an *OntologyAnnotation*.

*StringAnnotation* represents any string which a user can input as an annotation value, but values, represented by *OntologyAnnotation*, come from a language profile and, consequently, from an ontology.

While Multimedia Ontology is a crucial part of *OntoELAN*, a user can choose and load any domain ontology for annotation. In this paper, we use General Ontology for Linguistic Description (GOLD) for ontology-based annotation. In the next few paragraphs we give a primer on GOLD and refer the reader to [9] for details.

GOLD (available at http://www.u.arizona.edu/~farrar/gold.owl) is an ongoing research effort lead by the University of Arizona to define linguistic domain specific terms using OWL. GOLD contains four major categories of concepts:

- Expressions – physically accessible aspects of language (e.g. *WrittenLinguisticExpression*, *Word*, *WordPart*, *Prefix*).

- Grammar – the abstract properties and relations of language (e.g. *Tense*, *Number*, *Voice*).

- Data constructs – constructs that are used by linguists to analyze language data.

- Metaconcepts – the most basic concepts of linguistic analysis.

In our examples we will use GOLD concepts such as *Verb*, *Noun*, *Participle*, whose meanings are easy to understand without special training.

## 5. Language profile

A language profile is a subset of ontological terms, possibly renamed, that are used in the annotation of a particular multimedia resource. The idea of a language profile comes from the following two practical issues related to an ontology-based annotation.

First, a domain ontology defines all terms related to a particular domain, and the number of terms is usually considerably large. However, to annotate a concrete data resource, an annotator usually does not need all terms from an ontology. Moreover, an experienced annotator can identify a subset of ontological terms that will be useful for a given resource. Speaking in terms of a linguistic domain, an annotator will only use a subset of GOLD to annotate a particular language and may need a different subset for another language.

Second, linguists have been annotating multimedia data for years without standardized terms from an ontology. They have their individual sets of terms that they are accustomed to use for annotation. It will be difficult to come to a consensus about class names in GOLD, so that every linguist is satisfied with it. Additionally, linguists widely

```
<?xml version="1.0" encoding="UTF-8"?>
<PROFILE
    AUTHOR="Artem" DESCRIPTION="Potawatomi Language" VERSION="1.0"
    SOURCE="http://www.u.arizona.edu/~farrar/gold.owl">
  <USER_DEFINED_TERM DESCRIPTION="" NAME="NI">
    <ONTOLOGY_TERM NAME="Noun"/>
    <ONTOLOGY_TERM NAME="Inanimate"/>
  </USER_DEFINED_TERM>
</PROFILE>
```

**Figure 2. An example of a language profile XML document.**

use abbreviations like "n" for "noun" which is concise and convenient. Finally, linguists whose native language is, for example, Ukrainian may prefer to use annotation terms in Ukrainian rather than in English.

More formally, a language profile is defined as a quadruple: ontological terms; user-defined terms; a mapping between ontological terms and user-defined terms; a reference to an ontology, which contains the structural information about terms (like subclass relationship). A language profile in *OntoELAN* provides convenience and flexibility for a user to:

- select a subset of ontological terms useful for a particular resource annotation;

- rename ontological terms, e.g. use another language, give an abbreviation or a synonym;

- combine the meaning of two or many ontological terms in one user-defined term (e.g. ontological terms "Inanimate" and "Noun" may be conveniently renamed as "NI").

*OntoELAN* allows ontology-based annotation by means of a language profile. A user opens an ontology, creates a profile and links it to an ontological tier. Annotation values for an ontological tier can only be selected from a language profile.

A language profile in *OntoELAN* is represented as a simple XML document (see Figure 2) with a specified schema, which basically maps ontological terms to user-defined terms, has a link to an original ontology and some metadata. A user can easily create, open, edit and save profiles with *OntoELAN*.

Figure 2 presents a language profile, created by the author Artem and linked to GOLD ontology at URI http://www.u.arizona.edu/~farrar/gold.owl. In this example, there is only one user-defined term "NI" that maps to ontological terms "Noun" and "Inanimate". This is a one-to-many mapping, but, in general, a mapping is many-to-many. For example, we can add another user-defined term "IN" that maps to the same ontological terms "Noun" and "Inanimate". Obviously, a mapping can be one-to-one and many-to-one as well.

## 6. Annotation tiers and linguistic types

*OntoELAN* allows a user to create an unlimited number of annotation tiers. Multiple tier feature is a must for linguistic multimedia annotation. For example, while annotating an audio monolog, a linguist may choose separate tiers to write a monolog transcription, a translation, a part of speech annotation, a phonetic transcription, etc.

An annotation tier can be either *alignable* or *referring*. Alignable tiers are directly linked to the time axis of an audio/video clip and can be divided into segments (time slots); referring tiers contain annotations that are linked to annotation on another tier, which is also called *a parent tier* and can be alignable or referring. Thus, tiers can be viewed as a hierarchy, where its root must be an alignable tier. Following the previous example, the speech transcription could be an independent time-alignable tier which is divided into time slots of speaker's utterances. On the other hand, the translation referring tier could refer to the transcription tier, so that the translation tier inherits its time alignment from the transcription tier.

After a tier hierarchy is established, changes in one tier may influence other tiers. Deletion of a parent tier is cascaded: all its child tiers are automatically deleted. Similarly, this is true about annotations on a tier: deletion of an annotation on a parent tier causes the deletion of all corresponding annotations on its child tiers. Alteration of the time slot on a parent tier influences all child tiers also.

Each annotation tier has associated with it linguistic type. There are five predefined linguistic types in *OntoELAN*, which put some constraints on tiers assigned to them. The first four of them are described in [10], and we also cite their definitions here:

- *None*: the annotation on the tier is linked directly to the time axis. This is the only type that alignable tiers can have.

- *Time Subdivision*: the annotation on the parent tier can be subdivided into smaller units, which, in turn, can be linked to time slots. They differ from annotations on alignable tiers in that they are assigned to a slot that is contained within the slot of their parent annotation.

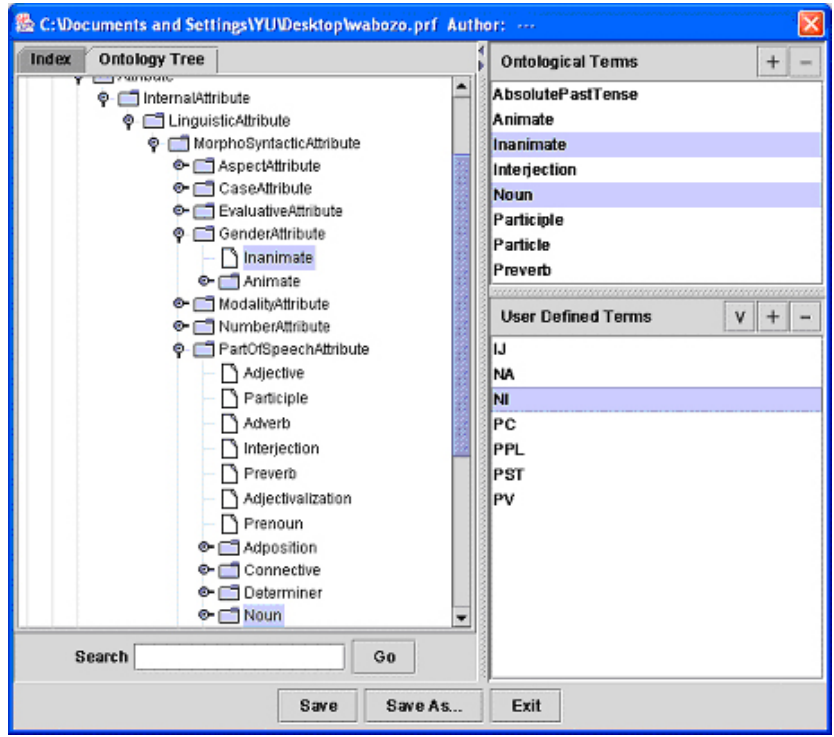

**Figure 3. A snapshot of creating the language profile.**

- *Symbolic Subdivision*: similar to the previous type, but the smaller units cannot be linked to the time slots.

- *Symbolic Association*: the annotation on the parent tier cannot be subdivided further, so there is a one-to-one correspondence between the parent annotation and its referring annotation.

- *Ontological Type*: the annotation on such a tier is linked to a language profile. This is not an independent type as it can be used only in combination with referring tier types such as *Time Subdivision*, *Symbolic Subdivision* or *Symbolic Association*. To emphasize that a referring tier allows ontology-based annotation, we call it an ontological tier.

Only ontological tiers allow annotation based on language profile terms; other types of tiers allow annotation with any string value.

## 7. Linguistic multimedia annotation with OntoELAN

In this section, we describe an annotation process in *OntoELAN* using a linguistic multimedia resource annotation example. In general, an annotation process in *OntoELAN* consists of three major steps: (1) language profile creation;

(2) creation of tiers; and (3) creation of annotations. The first step is unnecessary if ontological tiers will not be defined. The second step can be completed partially for non-ontological tiers before the creation of a language profile. It is also possible to have multiple profiles for multiple ontological tiers, but there is always one-to-one correspondence between a profile and an ontological tier.

As an example, we annotate the audio file, which contains a sentence in Potawatomi, one of the North American native languages.

We first load GOLD ontology and create the Potawatomi language profile. Figure 3 presents a snapshot of the profile creation window. Tabs "Index" and "Ontology Tree" on the left provide two views of an ontology: a list view, which displays all the terms of an ontology alphabetically as a list; a hierarchical view, which displays all the terms of an ontology in a hierarchical fashion to illustrate parent-child relationships between terms. From any of these two views, a user can select required terms and add them to the "Ontological Terms" list; rename ontological terms as shown in the "User Defined Terms" list. In Figure 3, we selected ontological terms "Inanimate" and "Noun" and combine them under one user-defined term "NI".

After the language profile is ready, we define six tiers in the *OntoELAN* main window (see Figure 4):

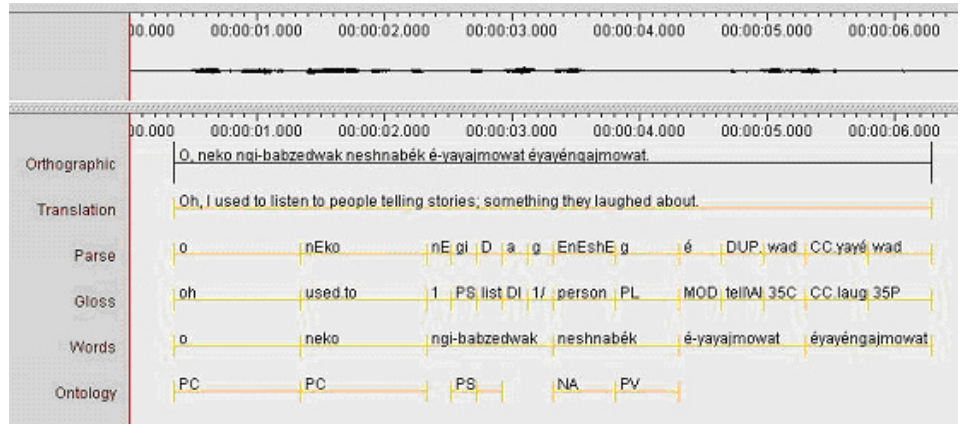

**Figure 4. A snapshot of annotation tiers in the OntoELAN main window.**

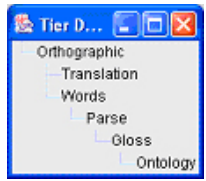

**Figure 5. A snapshot of the tier hierarchy.**

- Orthographic of type "None" (linked to the time axis);

- Translation of type "Symbolic Association" (referring to Orthographic);

- Words of type "Symbolic Subdivision" (referring to Orthographic);

- Parse of type "Symbolic Subdivision" (referring to Words);

- Gloss of type "Symbolic Association" (referring to Parse);

- Ontology of type "Symbolic Association" and "Ontological Type" (referring to Gloss).

The created tier hierarchy is shown in Figure 5.

Finally, we specify annotation values on all six tiers (see Figure 4). We annotate the *Orthographic* tier first, because it is the root of the tier hierarchy and its time alignment is inherited by other tiers. We do not divide *Orthographic* tier into time slots and its time axis contains the whole sentence in Potawatomi. *Translation* tier inherits time alignment from its parent and cannot subdivide it any further (type "Symbolic Association"). *Words* tier also inherits *Orthographic* time alignment, but in this case we subdivide it into segments that correspond to words in the sentence. Similarly, we subdivide *Parse* tier alignment inherited from *Words*. *Gloss* tier inherits alignment from *Parse*, and *Ontology* tier inherits alignment from *Gloss*; both *Gloss* and *Ontology* do not allow further subdivision. Correct alignment inheritance is important, because there is a semantic correspondence between segments of different tiers. For example, if we look at a Potawatomi word "neko" in *Words* tier, we can find its gloss "used to" in *Gloss* tier and part of speech "PC" (maps to GOLD *Participle* concept) in *Ontology* tier, etc.

Except for the annotations on the *Ontology* tier, which is defined as an ontological tier, all the annotations are annotated by a string value. Unlike the text annotation, the user annotates ontological tier by selecting a user-defined term from the profile. Once the term is selected, the next step is creating individuals of the corresponding ontological term(s). The user needs to do nothing but input an instance name if the ontological term is defined as an instance or a class with no restrictions. Otherwise, the user needs to create an instance of the ontological term based on the definition of the corresponding ontological class.

The annotation is saved in the XML format as instances of Multimedia Ontology and, in our case, GOLD. The example of the XML markup for the *Ontology* tier instance and referring annotation instance with ID "a42" on that tier is shown in Figure 6. For the *Ontology* tier, several properties are defined such as ID, parent tier, profile, linguistic type, etc. For the referring annotation, *OntoELAN* has defined ID, reference to another annotation, and annotation value that includes an *OntologyAnnotation* class instance with ID, user-defined term "PV" and reference to GOLD concept *Preverb*, which is defined as an instance. The markup in Figure 6 is based on Multimedia Ontology, except the reference to a GOLD instance mentioned above.

## 8. Conclusions and future work

We have developed *OntoELAN*, a linguistic multimedia annotation tool that features ontology-based annotation ap-

```
...
<media:Tier rdf:ID="Ontology">
<media:hasTierID>Ontology</media:hasTierID>
<media:hasParent rdf:resource="file:///C:/wabozo4.eaf#Gloss"/>
<media:hasProfile>C:\wabozo.prf</media:hasProfile>
<media:hasLinguisticType>
    <media:LinguisticType rdf:ID="ontology">
    <media:hasTimeAlignable>false</media:hasTimeAlignable>
    <media:hasLinguisticTypeID>ontology</media:hasLinguisticTypeID>
    <media:hasConstraint rdf:resource="file:///C:/wabozo4.eaf#Symbolic_Association"/>
    <media:hasGraphicRef>false</media:hasGraphicRef>
    </media:LinguisticType>
</media:hasLinguisticType>
...
</media:Tier>
...
<media:RefAnnotation rdf:ID="a42">
<media:hasAnnotationID>a42</media:hasAnnotationID>
<media:hasAnnotationRef rdf:resource="file:///C:/wabozo4.eaf#a31"/>
<media:hasAnnotationValue>
    <media:OntologyAnnotation rdf:ID="a42Value">
    <media:hasUserDefinedTerm>PV</media:hasUserDefinedTerm>
    <media:hasInstances
        rdf:resource="http://www.u.arizona.edu/~farrar/gold.owl#Preverb"/>
    <media:hasOntAnnotationDescription></media:hasOntAnnotationDescription>
    <media:hasOntAnnotationId>e</media:hasOntAnnotationId>
    </media:OntologyAnnotation>
</media:hasAnnotationValue>
</media:RefAnnotation>
...
```

**Figure 6. An example of the XML markup for the OntoELAN annotation.**

proach. *OntoELAN* is the first attempt at annotating linguistic multimedia data with a linguistic ontology. Meanwhile, the ontological annotations share the data on the linguistic ontologies. Future work may improve the system, providing more channels of sharing more data on the Web, such as the multimedia descriptions, the language words, etc. Also, a future version will improve the current searching system, which supports text searching and retrieve in one annotation document, to search, retrieve and compare the linguistic multimedia annotation data on the Web. Additionally, we plan to integrate a text document annotation into *OntoE-LAN* and include semi-automatic annotation support, similar to *Shoebox* [4].

## References

[1] IBM MPEG-7 Annotation Tool. *http://www.alphaworks.ibm.com/tech/videoannex*.

[2] NSF 04-605. Documenting Endangered Languages (DEL). *http://www.nsf.gov/pubs/2004/nsf04605/nsf04605.htm*.

[3] The Protégé Project. *http://protege.stanford.edu*.

[4] The Linguist's Shoebox: Tutorial and User's Guide. SIL International. *http://www.sil.org/computing/shoebox/*, 2000.

[5] S. Bechhofer, F. Harmelen, J. Hendler, I. Horrocks, D. McGuinness, P. Patel-Schneider, and L. Stein. OWL Web Ontology Language Reference. W3C Recommendation. *http://www.w3.org/TR/owl-ref/*, February 2004.

[6] T. Berners-Lee, J. Hendler, and O. Lassila. The Semantic Web. *Scientific American*, May 2001.

[7] T. Bray, J. Paoli, C. Sperberg-McQueen, E. Maler, and F. Yergeau. Extensible Markup Language (XML) 1.0 (Third Edition). W3C Recommendation. *http://www.w3.org/TR/REC-xml/*, February 2004.

[8] D. C. Fallside. XML Schema Part 0: Primer. W3C Recommendation. *http://www.w3.org/TR/xmlschema-0/*, May 2001.

[9] S. Farrar and D. T. Langendoen. A Linguistic Ontology for the Semantic Web. *GLOT International*, 7(3):97–100, 2003.

[10] B. Hellwig and D. V. Uytvanck. EUDICO Linguistic Annotator (ELAN) version 2.0.2 Manual. *http://www.mpi.nl/tools/ELAN/ELAN_Manual-04-04-08.pdf*.

[11] S. Lu, M. Dong, and F. Fotouhi. The Semantic Web: Opportunities and Challenges for Next-Generation Web Applications. *International Journal of Information Research*, 7(4), 2002.

[12] S. Lu, D. Liu, F. Fotouhi, M. Dong, R. Reynolds, A. Aristar, M. Ratliff, G. Nathan, J. Tan, and R. Powell. Language Engineering for the Semantic Web: a Digital Library for Endangered Languages. *International Journal of Information Research*, 9(3), April 2004.

[13] J. M. Martinez. MPEG-7 Overview (version 9). International Organisation for Standardisation, ISO/IEC JTC1/SC29/WG11, Coding of Moving Pictures and Audio. *http://www.chiariglione.org/mpeg/standards/mpeg-7/mpeg-7.htm*, March 2003.